\documentclass{aa}
\usepackage[varg]{txfonts}

\usepackage{graphicx}
\usepackage{tikz}
\usepackage[breaklinks=true]{hyperref}
\bibpunct{(}{)}{;}{a}{}{,} 

\newcommand{\thru}{\ensuremath{\text{--}}}

\newcommand{\rhel}{\ensuremath{r_\mathrm{h}}}

\newcommand{\um}{\ensuremath{\,\mathrm{\mu m}}}

\newcommand{\km}{\ensuremath{\,\mathrm{km}}}

\newcommand{\polr}{\ensuremath{P_\mathrm{r}}}
\newcommand{\thr}{\ensuremath{\theta_\mathrm{r}}}

\newcommand{\Pmin}{\ensuremath{P_\mathrm{min}}}
\newcommand{\Pmax}{\ensuremath{P_\mathrm{max}}}
\newcommand{\amin}{\ensuremath{\alpha_\mathrm{min}}}

\newcommand{\ainv}{\ensuremath{\alpha_0}}
\newcommand{\pp}{\ensuremath{\,\mathrm{\%p}}}
\newcommand{\Ks}{\ensuremath{\mathrm{K_s}}}

\newcommand{\PaperI}{Paper I}

\begin{document}

\title{Quantitative grain size estimation on airless bodies \\ from the negative polarization branch}
\subtitle{II. Dawn mission targets (4) Vesta and (1) Ceres}

\author{
  Yoonsoo P. Bach\inst{1, 2, 3}
  \and Masateru Ishiguro\inst{2, 3}
  \and Jun Takahashi \inst{4}
  \and Jooyeon Geem\inst{2, 3}
  \and \\ Daisuke Kuroda \inst{5}
  \and Hiroyuki Naito \inst{6}
  \and Jungmi Kwon \inst{7}
}

\institute{
  Korea Astronomy and Space Science Institute (KASI), 776 Daedeok-daero, Yuseong-gu, Daejeon 34055, Republic of Korea
  \and
  Department of Physics and Astronomy, Seoul National University, Gwanak-ro 1, Gwanak-gu, Seoul 08826, Republic of Korea
  \and
  SNU Astronomy Research Center, Department of Physics and Astronomy, Seoul National University, Gwanak-ro 1, Gwanak-gu, Seoul 08826, Republic of Korea
  \and
  Center for Astronomy, University of Hyogo, 407-2 Nishigaichi, Sayo, Hyogo 679-5313, Japan
  \and
  Japan Spaceguard Association, Bisei Spaceguard Center 1716-3 Okura, Bisei, Ibara, Okayama 714-1411, Japan
  \and
  Nayoro Observatory, 157-1 Nisshin, Nayoro, Hokkaido 096-0066, Japan
  \and
  Department of Astronomy, Graduate School of Science, The University of Tokyo, 7-3-1 Hongo, Bunkyo-ku, Tokyo 113-0033, Japan
  \\ \email{ysbach93@gmail.com, ishiguro@snu.ac.kr}}

\date{Received ; accepted }

\abstract
{Sunlight scattered from the surface of an airless body is generally partially polarized, and the  corresponding polarization state includes information about the scattering surface, such as albedo, surface grain sizes, composition, and taxonomic types.
Most polarimetric studies on airless bodies thus far have focused on optical wavelengths ($ \lambda \lesssim 1 \,\mathrm{\mu m} $).}
{We conducted polarimetry of two large airless bodies, the Dawn mission targets (1) Ceres and (4) Vesta, in the near-infrared region. We further investigated the change in the polarimetric phase curves over the wavelengths expected from previous works.
}
{We used the Nishiharima Infrared Camera (NIC) installed at the Nishi-Harima Astronomical Observatory (NHAO) to observe these objects at multiple geometric configurations in the J, H, and \Ks\ bands ($ \lambda \sim 1.2\mathrm{-}2.3 \,\mathrm{\mu m} $).}
{Polarimetric parameters were determined and compared with previously reported experimental results. In particular, Vesta exhibits a characteristic change in the negative polarization branch as the wavelength increases to the \Ks\ band, which we interpret as an indication of the dominant existence of $D \sim 10\thru20\um$ particles. Our approach is supported by empirical reasoning and coincides well with an independent, theory-driven approach based on thermal modeling.}
{This work demonstrates how near-infrared polarimetry can be utilized to quantitatively determine the particle size of airless objects. This finding will have important implications for asteroid taxonomy and regolith evolution.}


\titlerunning{NIR polarimetry of airless bodies}
\maketitle

\section{Introduction}
The presence of a regolith on an airless body is the result of its initial physical and chemical conditions and specific history of evolution. Among the parameters used to characterize the regolith are grain size, which is related to thermal parameters \citep{2013Icar..223..479G}, and scattering processes \citep{Hapke2012}, which can change the observable parameters. Thus, quantification of these parameters is one of the important steps in modeling observed quantities. However, the fine nature of grains, as seen with the submicron-scale fine particles retrieved from the Moon \citep{2008ParkJ+Lunar}, makes it difficult to quantify their sizes, even in this space exploration era.

In our previous work (Bach et al. 2024; {\PaperI} hereafter), our extensive investigation established a methodology for quantifying grain sizes on airless bodies through multiwavelength polarimetry. We focused on exploring the widening-and-deepening (WD) trend within the negative polarization branch (NPB) as triggered by changes in albedo, $A(\lambda)$, and the size parameter, $X \propto D/\lambda$, where $D$ represents grain size and $\lambda$ denotes the observation wavelength. We identified two significant effects influencing the NPB across wavelengths. First, as $A$ increases, we observe a WD-like trend for $A\lesssim 10\%$ and a shallowing trend for $A \gtrsim 10\%$. Second, regarding $D/\lambda$, we found that the NPB stays relatively unchanged for $D/\lambda \gg 10$. However, as $\lambda$ increases or $D/\lambda$ decreases, the WD trend starts to appear at $D/\lambda \lesssim 5\thru10$. By focusing on the effect of $D/\lambda$, we can estimate the particle size that contributes to the scattering process.

Although it is a useful approach, interpreting the WD trend remains challenging due to the potential for a convoluted influence of both albedo and $D/\lambda$. Thus, observations of (i) airless objects (ii) covered with fine grains and (iii) exhibiting relatively constant reflectance across visible to near-infrared (NIR) wavelengths are necessary for more reliable interpretations when focusing on the $D/\lambda$ effect for cases free of the effect of albedo.


The particle size on asteroids is roughly related to the size of the parent asteroid; the larger the asteroid is, the smaller the particle size (\citealt{2007Icar..190..236D}; however, see \citealt{2022PSJ.....3...47M,2022PSJ.....3..263M}). This is generally understood to be due to the stronger gravity of larger objects \citep{2014GeoRL..41.1438C} and/or the shorter lifetime of smaller objects \citep{2015aste.book..107D}.
For (1) Ceres (size $d \sim 950 \km$), the presence of ultra-porous fine particles ($D \ll 1 \um$) may be required to explain the observations (\citealt{2019Icar..322..144L, 2021NatCo..12..274S}; Sect. \ref{ss:case of ceres}). Similarly, spectroscopic analyses suggest that (4) Vesta ($d \sim 530 \km$) is predominantly covered with particles of sizes $\lesssim 20-25 \um$ \citep{1994Metic..29..394H, 2011Icar..216..640L, 2019MNRAS.483.1952M}, and optical polarimetry has indicated the possibility of a mixture of eucrite grains with sizes $> 74 \um$ and $< 10\um$ \citealt{1980Icar...43..172L}; also see Sect. \ref{ss:vesta particle size}).

Therefore, in our study we chose to investigate Ceres and Vesta because (i) they are among the brightest airless objects, (ii) they are likely to possess fine particles that can display the $D/\lambda$ effect (\PaperI), (iii) both have reflectances that suggest they are not subject to a strong albedo effect  (\PaperI), and (iv) they are targets of the Dawn mission, so a large amount of information is available.

Our NIR polarimetry investigation of airless bodies appears to have commenced nearly simultaneously with that described in \cite{2022PSJ.....3...90M}, where the first observations in the J and H bands from UT 2019-03-17 are reported. To the best of our knowledge, \cite{2022PSJ.....3...90M} is the first peer-reviewed publication of NIR polarimetry of airless bodies; it includes our own observations that have been conducted since UT 2019-10-22, at which time we were unaware of their pioneering work. In their work, the albedo and refractive indices, but not the grain size of asteroids, are discussed. In a later study \citep{2023PSJ.....4...93M}, they proposed a qualitative description of the existence of fine-grained inclusions on Barbarian asteroids. Both publications studied wavelengths up to the H band ($\lambda \sim 1.6 \um$). We stress that our study is the first NIR polarimetry study of a V-type asteroid (Vesta) and the first polarimetry study of airless bodies at $\lambda > 2\um$ (the \Ks \ band), a range that contains critical information (Sect. \ref{ss:vesta particle size}).

In this study we conducted a comprehensive investigation utilizing our new data and previous experimental results. The observational circumstances, data reduction methods, and polarimetric parameter finding methods used are summarized in Sect. \ref{s: method}. The results of our observations are presented in Sect. \ref{s: result}. In Sect. \ref{s:disc} we discuss how our results can be interpreted and how the grain sizes of these targets can be obtained. Finally, the appendix provides details on our data reduction and statistical analyses as well as information regarding the data and code availability.


\section{Observations and methods} \label{s: method}

\subsection{Instruments and observations}
Our observations were conducted using the Nishiharima Infrared Camera (NIC) attached to the 2.0-m Nayuta telescope at the Nishi-Harima Astronomical Observatory (NHAO; 134.3356$ ^\circ $E, 35.0253$ ^\circ $ N, 449 m elevation). The telescope has an altazimuth mount with a f/12.0 focal ratio. The system consists of three detector arrays, each $1024 \times 1024$ pixel, with two dichroic mirrors such that the three wavelengths are imaged simultaneously. The observed wavelengths are
J (central wavelength $ \lambda = 1.253 \um $ with a full width at half maximum $ \Delta \lambda = 0.157 \um $),
H ($ \lambda = 1.632 \um$, $ \Delta \lambda = 0.298 \um $), and
\Ks ($ \lambda = 2.146 \um $, $ \Delta \lambda = 0.313 \um$).

In the optical path, a polarimetric mask, a rotating half-wave plate (HWP), and a beam displacer are configured \citep{TakahashiJ2018SAG, TakahashiJ2019SAG}. As a result, two images, ordinary and extraordinary rays (hereafter, o- and e-rays), appear on the same detector at a pixel scale of $ 0.16 \arcsec/\mathrm{pix} $. Each image has a size of $\sim 150 \times 430 \,\mathrm{pixels}$ ($ 24\arcsec \times 69\arcsec $) and is displaced horizontally by approximately 200 pixels. Due to the beam displacer, the point spread functions are expected to be different for o- and e-rays

Each set of polarimetric observations consists of four exposures for which with the HWP angle was set to $ 0^\circ $, $ 45^\circ $, $ 22.5^\circ $, and $ 67.5^\circ $, respectively. The four exposures together are denoted as a ``set,'' with  the first two and last two exposures denoted as the $ q $ and $ u $ subsets, respectively. These parameters are used to determine the normalized Stokes $ q = Q/I $ and $ u = U/I $ parameters, respectively. We conducted five- and six-night polarimetric observations for Ceres and Vesta, respectively. The observational conditions are summarized in Table \ref{tab: obs}. On UT2019-11-08, when Vesta was at $\alpha=4.1\degr$, the weather conditions were very unfavorable. This circumstance led to widely varying signal-to-noise ratios (S/Ns) even between consecutive exposures. Consequently, the uncertainty, particularly in the H-band measurements, was substantially elevated during this period (Table \ref{tab: res-pol}).

\begin{table*}[!tb]
\caption[Observing circumstances.]{Observing circumstances. }
\label{tab: obs}
\centering
\begin{tabular}{lcc|cccccl}
\hline \hline
Date & Time & Object & $N$ & EXPTIME & $\rhel$ & $r_\mathrm{o}$ & $\alpha$ & Skymonitor record\\
 & UT &  &  & sec & au & au & ° & \\
\hline
2019-10-22  & 15:27-16:04 & Vesta & 21 & 2 & 2.5 & 1.6 & 10.4 & \url{https://youtu.be/PaU3ign_sdY}\\
2019-11-08\tablefootmark{(a)} & 12:24-17:27 & Vesta & 145 & 2, 3, 4, 5, 10, 30 & 2.5 & 1.6 & 4.1 & \url{https://youtu.be/dg5ejH7fPd8}\\
2019-11-21 & 12:22-16:37 & Vesta & 165 & 2 & 2.6 & 1.6 & 5.7 & \url{https://youtu.be/_gHz8hH7eXQ}\\
2019-12-18 & 13:20-14:57 & Vesta & 67 & 2, 3 & 2.6 & 1.8 & 15.6 & \url{https://youtu.be/ceSwHR12S-s}\\
2020-01-10 & 13:18-13:59 & Vesta & 31 & 2 & 2.6 & 2.0 & 20.7 & \url{https://youtu.be/3Wkn83F4Qzg}\\
2020-02-13 & 09:04-10:00 & Vesta & 30 & 2 & 2.6 & 2.5 & 22.5 & \url{https://youtu.be/VzwG5y-18ME}\\
\hline
2020-06-21  & 18:30-19:20 & Ceres & 18 & 20 & 3.0 & 2.5 & 18.9 & \url{https://youtu.be/Y4H6QNyWcp0} \\
2020-10-02\tablefootmark{(b)} & 09:19-10:30 & Ceres & 20 & 20 & 3.0 & 2.2 & 13.1 & \url{https://youtu.be/muyaqzYcnb4} \\
2020-10-18 & 08:50-09:59 & Ceres & 22 & 20 & 3.0 & 2.3 & 16.5 & \url{https://youtu.be/6-SD0WrY1No} \\
2021-09-09\tablefootmark{(c)} & 16:15-17:45 & Ceres & 25 & 20 & 2.8 & 2.5 & 20.8 & \url{https://youtu.be/fUuu9HiQT8M} \\
2021-11-07 & 12:34-22:40 & Ceres & 31 & 5 & 2.8 & 1.8 & 8.6 & \url{https://youtu.be/W7xduRk5tSM} \\
\hline
\end{tabular}
\tablefoot{
 Date: The UT date in YYYY-MM-DD. Time: The starting and ending UT time in HH:mm. $N$: The total number of obtained polarimetric sets (one set consists of four exposures). EXPTIME: The exposure times used. $\rhel$: The heliocentric distance to the target. $r_\mathrm{o}$: The observer-centric distance to the target.\   $\alpha$: The phase angle. Skymonitor record: The NHAO skymonitor camera record archived on YouTube (also available in Appendix \ref{s:data code avail}).
  \tablefoottext{a}{Unstable weather (refer to the skymonitor record).}
  \tablefoottext{b}{Most H-band frames are saturated.}
  \tablefoottext{c}{All H-band frames are saturated.}
}
\end{table*}

\subsection{Data reduction process}
The data reduction process follows a similar approach to that of our previous publication (Appendix B of \citealt{2021A&A...653A..99T}), with some modifications implemented using the \texttt{NICPolpy} software\footnote{
https://github.com/ysBach/NICpolpy, https://pypi.org/project/NICpolpy/
} \citep{2022_SAG_NICpolpy}. The instrumental gain and readout noise parameters were also updated from \cite{Ishiguro2011ARNHAO}. Two polarimetric dome flats were used, one taken on UT 2018-05-07 and the other taken on UT 2020-06-03. For data taken after June 2020, the second master flat was utilized.

The preprocessing steps are as follows: First, artifacts such as vertical and Fourier-like patterns in the frames are removed. Then, standard dark subtraction and flat-fielding are applied. The values at bad pixels, identified using a pre-made bad pixel mask, are replaced with interpolated values from nearby pixels. Each frame is split into two FITS files, which represent the o-ray and e-ray images. Finally, cosmic-ray rejection is performed. Detailed explanations and the implemented algorithms are presented in \cite{2022_SAG_NICpolpy}.

After preprocessing, signal extraction was performed for each exposure. The Stokes parameters were estimated using the ratio method \citep[see, e.g.,][]{2009PASP..121..993B}, considering each subset of exposures. To ensure the robustness of our polarization degree estimation, we extensively tested our results against multiple signal extraction techniques and outlier rejection schemes (see Appendix \ref{s: polarimetry detail}). We also conducted thorough verifications to confirm the stability of our final results against different parameter values used in preprocessing, such as those related to artifact removal, the use of different flats, bad-pixel interpolation, and cosmic-ray rejection parameters.
These methods only marginally changed the intermediate results, such that the final results (i.e., those describing the polarimetric parameters) do not vary beyond the measurement uncertainty range.

\subsection{Polarization parameter calculation}
The polarization degree of solar system objects is often described by the properly rotated polarization degree \citep{1929PhDT.........9L}:
\begin{equation}\label{eq: polr}
  \polr \coloneqq P \cos (2 \thr) \approx \frac{I_\perp - I_\parallel}{I_\perp + I_\parallel}~,
\end{equation}
where $P$ is the total linear polarization degree, $\thr$ is the position angle of the strongest electric vector relative to the normal vector of the scattering plane (e.g., $0$ or $180^\circ$ indicates it is normal to the scattering plane), and $I_\perp$ and $I_\parallel$ denote the measured intensities of scattered light along the directions perpendicular and parallel to the scattering plane, respectively.

The curve describing this phenomenon as a function of the phase angle (light source-target-observer angle; $\alpha \coloneqq 180\degr - \mathrm{scattering~angle}$), $\polr(\alpha)$, is denoted the polarization phase curve (PPC). A PPC at small $\alpha$ can be expressed with a three-parameter linear-exponential function \citep{2002MmSAI..73..716M,2003Icar..161...34K}:
\begin{equation}\label{eq: linexp original}
  \polr(\alpha) = \theta_1 \left ( e^{-\alpha/\theta_2} - 1 \right ) + \theta_3 \alpha ~,
\end{equation}
for free parameters $\theta_{1,\, 2,\, 3}$. The inversion angle $\ainv$ can be solved numerically from $\polr(\alpha_0) = 0$, as in \cite{gil-hutton-data}.

We emphasize that the three parameters used above are eventually converted to polarimetric parameters of interest, for example the polarimetric slope $h \coloneqq \left . \frac{d\polr}{d\alpha} \right |_{\alpha=\ainv} = \polr^\prime(\alpha=\alpha_0)$, inversion angle $\ainv$, $\amin$ and $\Pmin = \polr(\amin)$. By rearranging the equation, we can generate the fitting parameters with more intuitive meanings and facilitate subsequent statistical analyses of the parameters:
\begin{equation}\label{eq: linexp}
  \polr(\alpha) = \polr(\alpha; h,\, \ainv,\, k)
    = h\frac{(1 - e^{-\alpha_0/k})\alpha - (1 - e^{-\alpha/k})\alpha_0}{1 - (1 + \alpha_0/k) e^{-\alpha_0/k}} ~,
\end{equation}
where $k \, [\degr]$ is a scale parameter. Having rearranged the equations such that two out of the three fitting parameters ($h$, $\ainv$) are the ones of interest, the error analyses become much more straightforward. Importantly, the function in Eq. (\ref{eq: linexp}) describes the identical curve as in Eq. (\ref{eq: linexp original}). Furthermore, we can obtain $\amin$ through the condition $\polr^\prime(\alpha=\amin) = 0$ as follows:
\begin{equation}\label{eq: amin}
  \amin = -k \ln \left \{ \frac{k}{\alpha_0} \left ( 1 - e^{-\alpha_0/k} \right ) \right \} ~.
\end{equation}
Therefore, four important parameters, $h$, $\ainv$, $\amin$, and $\Pmin = \polr(\amin)$, are obtained.

\section{Results} \label{s: result}

\subsection{PPC of targets}\label{ss:ppc of targets}
The polarization degrees derived from our data are summarized in Table \ref{tab: res-pol}. These polarization degrees ($P$) are obtained using the formula $P = \sqrt{q^2 + u^2}$, where $q$ and $u$ are the instrumental polarization efficiency and the position-angle corrected values, respectively \citep{TakahashiJ2019SAG}.
To calculate the uncertainties in $q$ and $u$, we first computed the weighted average and its uncertainty, $(\sum \sigma_i^{-2})^{-1/2} $, where $\sigma_i$ is the uncertainty of the $i$-th data point. The final uncertainties, $dq$ and $du$, were determined as the larger value recorded between the weighted average uncertainty and the standard error (i.e., the sample standard deviation divided by the square root of the number of samples). Because the position angle of the scattering plane at the time of observation has negligible uncertainty, $d\theta_P = d\thr$. For $d\polr$, we applied error propagation to Equation (\ref{eq: polr}), while neglecting the covariance between $dP$ and $d\thr$. The maximum value between this propagated uncertainty and $dP$ is obtained as $d\polr$.

\begin{table*}
\caption{Polarimetric data reduction results.}
\label{tab: res-pol}
\centering
\begin{tabular}{lccccccccc}
\hline\hline
Object & Filter & $n$ & $\alpha$ & $P$ & $\theta_P$ & $\polr$ & $d\polr$ & $\theta_\mathrm{r}$ & $d\theta_\mathrm{r}$ \\
  & & & $\degr$ & \% & $\degr$ & \% & \% & $\degr$ & $\degr$ \\
\hline
\textit{Ceres}\tablefootmark{(a)} & \textit{J}\tablefootmark{(a)} & $ - $ & $ 8.3$ & $  -  $ & $   -  $ & $-1.740$\tablefootmark{(a)} & $0.020$\tablefootmark{(a)} & $ 83.40$\tablefootmark{(a)} & $ 0.10$\tablefootmark{(a)} \\
\textit{Ceres}\tablefootmark{(a)} & \textit{J}\tablefootmark{(a)} & $ - $ & $21.0$ & $  -  $ & $   -  $ & $+0.840$\tablefootmark{(a)} & $0.020$\tablefootmark{(a)} & $ 22.30$\tablefootmark{(a)} & $ 0.50$\tablefootmark{(a)} \\
Ceres & J   &  31 & $ 8.6$ & $1.620$ & $ -2.37$ & $-1.619$ & $0.038$ & $ 88.91$ & $ 0.67$ \\
Ceres & J   &  17 & $13.1$ & $1.054$ & $+39.70$ & $-1.053$ & $0.023$ & $ 88.97$ & $ 0.61$ \\
Ceres & J   &  14 & $16.5$ & $0.388$ & $+35.13$ & $-0.383$ & $0.029$ & $ 85.44$ & $ 2.11$ \\
Ceres & J   &  17 & $18.9$ & $0.289$ & $-68.86$ & $+0.288$ & $0.032$ & $  1.58$ & $ 3.15$ \\
Ceres & J   &  25 & $20.8$ & $0.898$ & $-83.42$ & $+0.898$ & $0.031$ & $  1.01$ & $ 1.00$ \\
\hline
\textit{Ceres}\tablefootmark{(a)} & \textit{H}\tablefootmark{(a)} & $ - $ & $ 8.3$ & $  -  $ & $   -  $ & $-1.610$\tablefootmark{(a)} & $0.030$\tablefootmark{(a)} & $ 86.60$\tablefootmark{(a)} & $ 0.10$\tablefootmark{(a)} \\
\textit{Ceres}\tablefootmark{(a)} & \textit{H}\tablefootmark{(a)} & $ - $ & $21.0$ & $  -  $ & $   -  $ & $+0.880$\tablefootmark{(a)} & $0.020$\tablefootmark{(a)} & $ 16.20$\tablefootmark{(a)} & $ 0.20$\tablefootmark{(a)} \\
Ceres & H   &  31 & $ 8.6$ & $1.558$ & $ -2.82$ & $-1.558$ & $0.046$ & $ 89.36$ & $ 0.85$ \\
Ceres & H   &   4 & $13.1$ & $1.155$ & $+40.00$ & $-1.153$ & $0.037$ & $ 88.68$ & $ 0.92$ \\
Ceres & H   &  15 & $16.5$ & $0.273$ & $+32.77$ & $-0.273$ & $0.028$ & $ 87.80$ & $ 2.97$ \\
Ceres & H   &  17 & $18.9$ & $0.356$ & $-69.82$ & $+0.356$ & $0.029$ & $  0.61$ & $ 2.35$ \\
\hline
Ceres & \Ks &  31 & $ 8.6$ & $1.626$ & $ -1.92$ & $-1.624$ & $0.039$ & $ 88.46$ & $ 0.68$ \\
Ceres & \Ks &  20 & $13.1$ & $1.063$ & $+40.65$ & $-1.061$ & $0.049$ & $ 88.03$ & $ 1.33$ \\
Ceres & \Ks &  14 & $16.5$ & $0.358$ & $+32.52$ & $-0.357$ & $0.046$ & $ 88.05$ & $ 3.72$ \\
Ceres & \Ks &  13 & $18.9$ & $0.535$ & $-70.83$ & $+0.535$ & $0.050$ & $  0.40$ & $ 2.69$ \\
Ceres & \Ks &  24 & $20.8$ & $1.105$ & $-80.27$ & $+1.102$ & $0.041$ & $  2.14$ & $ 1.06$ \\
\hline
Vesta & J   &  55 & $ 4.1$ & $0.657$ & $-49.51$ & $-0.657$ & $0.038$ & $ 89.53$ & $ 1.63$ \\
Vesta & J   & 158 & $ 5.7$ & $0.659$ & $+52.90$ & $-0.659$ & $0.028$ & $ 89.42$ & $ 1.22$ \\
Vesta & J   &  20 & $10.4$ & $0.674$ & $ -7.78$ & $-0.672$ & $0.089$ & $ 87.92$ & $ 3.77$ \\
Vesta & J   &  67 & $15.6$ & $0.413$ & $+22.18$ & $-0.411$ & $0.033$ & $ 87.13$ & $ 2.28$ \\
Vesta & J   &  31 & $20.7$ & $0.128$ & $+11.06$ & $-0.122$ & $0.033$ & $ 81.19$ & $ 7.43$ \\
Vesta & J   &  30 & $22.5$ & $0.069$ & $-28.91$ & $-0.002$ & $0.047$ & $ 45.69$ & $19.78$ \\
\hline
Vesta & H   &  16 & $ 4.1$ & $0.636$ & $-52.04$ & $-0.632$ & $0.255$ & $ 87.05$ & $11.48$ \\
Vesta & H   &  99 & $ 5.7$ & $0.773$ & $+52.92$ & $-0.773$ & $0.037$ & $ 89.48$ & $ 1.36$ \\
Vesta & H   &  18 & $10.4$ & $0.792$ & $ +1.05$ & $-0.770$ & $0.078$ & $ 83.25$ & $ 2.82$ \\
Vesta & H   &  67 & $15.6$ & $0.582$ & $+21.48$ & $-0.578$ & $0.033$ & $ 86.43$ & $ 1.60$ \\
Vesta & H   &  31 & $20.7$ & $0.197$ & $+15.94$ & $-0.195$ & $0.033$ & $ 86.08$ & $ 4.77$ \\
Vesta & H   &  30 & $22.5$ & $0.077$ & $ +4.72$ & $-0.072$ & $0.041$ & $ 79.32$ & $15.30$ \\
\hline
Vesta & \Ks &  45 & $ 4.1$ & $0.906$ & $-49.01$ & $-0.906$ & $0.048$ & $ 89.97$ & $ 1.51$ \\
Vesta & \Ks & 158 & $ 5.7$ & $0.984$ & $+51.64$ & $-0.982$ & $0.047$ & $ 88.16$ & $ 1.37$ \\
Vesta & \Ks &  20 & $10.4$ & $1.017$ & $ -1.04$ & $-1.003$ & $0.093$ & $ 85.34$ & $ 2.63$ \\
Vesta & \Ks &  66 & $15.6$ & $0.710$ & $+20.74$ & $-0.702$ & $0.037$ & $ 85.69$ & $ 1.51$ \\
Vesta & \Ks &  29 & $20.7$ & $0.366$ & $ +8.43$ & $-0.337$ & $0.044$ & $ 78.56$ & $ 3.46$ \\
Vesta & \Ks &  29 & $22.5$ & $0.160$ & $-15.27$ & $-0.077$ & $0.055$ & $ 59.33$ & $ 9.81$ \\
\hline
\end{tabular}
\tablefoot{
  $n$: Number of used polarimetric datasets after data filtering and outlier rejections (one set corresponds to four exposures). $\alpha$: Solar phase angle. $P$ and $dP$: The total linear polarization degree and its uncertainty. $\theta_P$: The position angle of the strongest electric field vector. $\polr$ and $d\polr$: The proper linear polarization degree referring to the scattering plan and its uncertainty.  $\thr$ and $d\thr$: Position angle of the strongest electric vector, with respect to the scattering plane normal vector, and its uncertainty. See Sect. \ref{ss:ppc of targets} for the details of error calculations.
  \tablefoottext{a}{Data from \cite{2022PSJ.....3...90M}.}
}
\end{table*}

Notably, these uncertainties are comparable to the standard error from the central limit theorem, which is inherently much less than the random scatter (standard deviation of data points or the error bar of a single data point) by a factor of $\approx \sqrt{n}$. In their work, \cite{2023PSJ.....4...93M} accounted for this discrepancy by quadratically adding $0.1\pp$ scatter to their reported $d\polr$ values. Additionally, these uncertainties do not include systematic errors (see Appendix \ref{ss:systematics}).

As shown in Table \ref{tab: res-pol}, we have confirmed that our $\thr$ values are generally close to integer multiples of $90\degr$, as expected from symmetry considerations \citep{1929PhDT.........9L, 2018MNRAS.481L..49C}. Consequently, $\cos(2\thr) \approx 1$. The extreme exception is Vesta at $\alpha = 22.5\degr$, where its polarization degree is near zero; thus, $\thr$ is not well defined in this particular case. This finding strongly supports the credibility of our data reduction process. Conversely, we can further utilize our $\thr$ values of asteroids to measure the instrumental offset (assuming it must be an integer multiple of $90\degr$) when $|\polr|$ is large. For example, using Ceres data at $\alpha=8.6\degr$, we find that the instrumental position angle offset is $< 2 \degr$ for all the filters, which is consistent with \cite{TakahashiJ2019SAG}.

\begin{figure*}[!tb]
  \centering
  \includegraphics[width=0.95\linewidth]{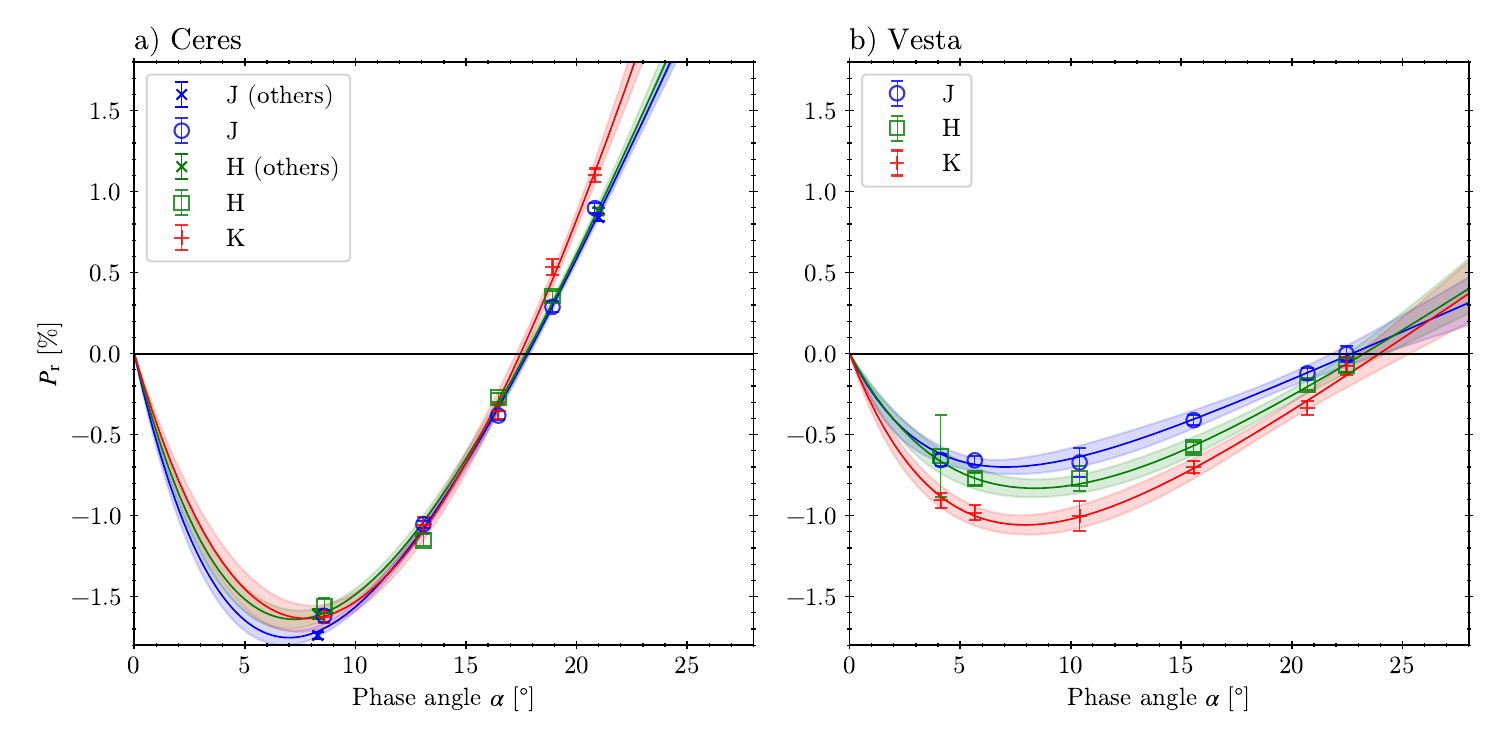}
\caption{PPCs of two objects in the NIR data from this study (star) and a previous publication (circle; \citealt{2022PSJ.....3...90M}). The solid lines are the least-square solutions, and the shaded areas show the MC 1$\sigma$ bounds.}
  \label{fig:ppc}
\end{figure*}

The PPCs for Ceres and Vesta are derived and plotted in Fig. \ref{fig:ppc}.
For Ceres, two datasets from a previous publication were included \citep{2022PSJ.....3...90M}. We have overplotted these data in the figures, and the results of our observations closely match the corresponding fitting curves.
The shaded regions in the figures represent the approximate 1$\sigma$ range of the fitting models obtained from Monte Carlo (MC) simulations (Appendix \ref{s: mcmc}).

\begin{figure*}[tb]
  \centering
  \includegraphics[width=0.95\linewidth]{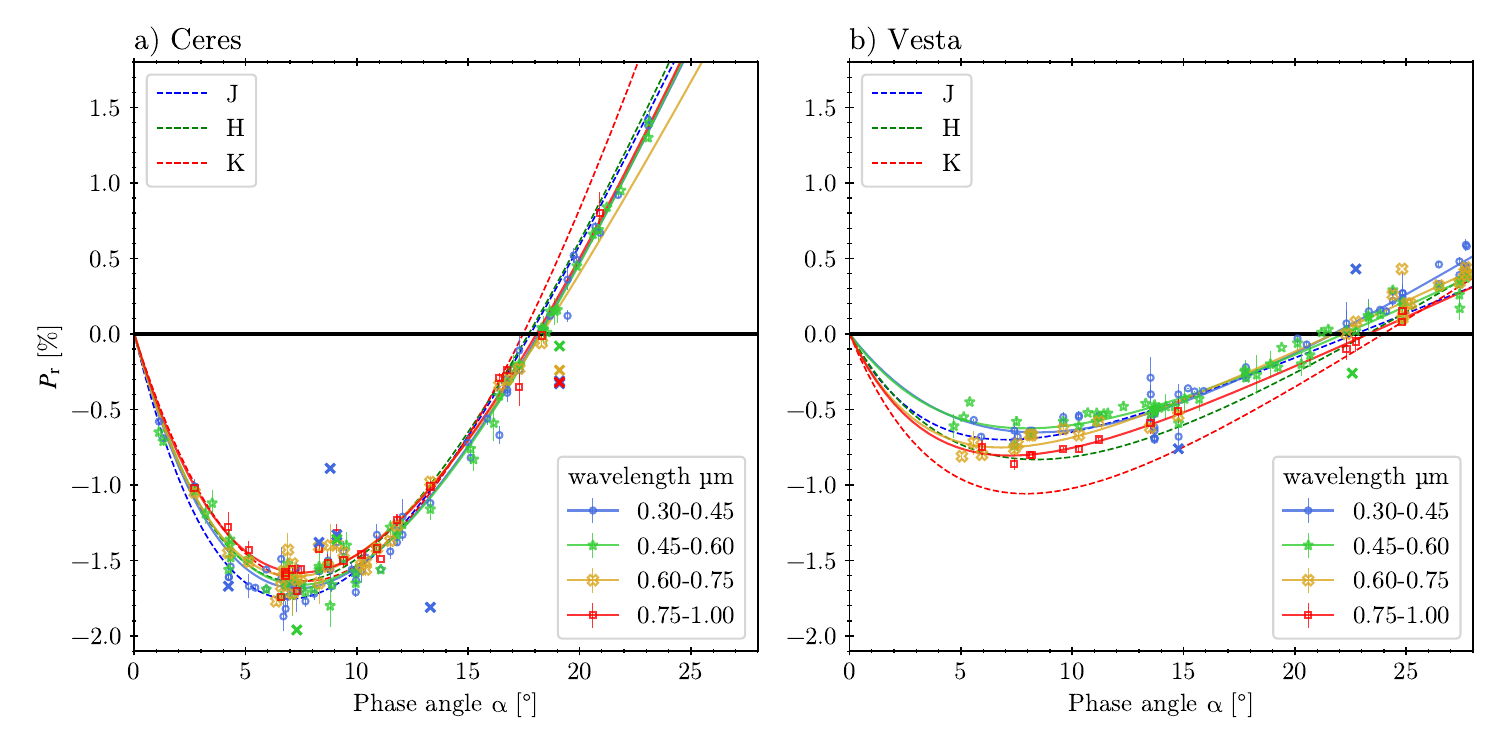}
\caption{Same as Fig. \ref{fig:ppc} but with optical data \citep{2022pdss.data....1L}. The solid lines are the least squares solutions for each wavelength range of the data, and the cross markers  (``x'') indicate the masked data points.}
  \label{fig:ppc+apd}
\end{figure*}

In Figure \ref{fig:ppc+apd}, we present the same data and curves as shown above, but with the inclusion of optical observations from \cite{2022pdss.data....1L}\footnote{
We used $\polr=-P$ if neither $\polr$ nor $\thr$ was available since such data are from $\alpha \ll \ainv$. We did not use R2 filter ($1/\lambda = 1.20 \um^{-1}$) data (2 observations for both targets were adopted from \citealt{1974AJ.....79.1100Z}) because they are largely scattered from other data in the I filter ($1/\lambda = 1.05 \um^{-1}$). Additionally, two data points in the $10\um$ N band \citep{1983Icar...56..381J} were mistakenly labeled as the $0.333\um$ N band; therefore, we did not use these data.
} and \cite{2022A&A...665A..66B}. For Vesta, we also used the data provided in \cite{2016MNRAS.456..248C}, which were not included in previous datasets (at eight different $\alpha$s), after taking the nightly median (3-sigma-clipped). When fitting the optical data, we used data only when the reported error bar for the polarization degree was $<0.15\pp$; we neglected any uncertainties in the data because of the inhomogeneity in the data across multiple sources. Notably, we expect optical data to have higher uncertainties. For example, a few noisy Ceres data points near $\ainv$ affect the $h$ values (i.e., they have high leverage values). Finally, we excluded data points with $>0.25\pp$ deviation from the fitting line.
We explored various processing methods, such as using only the data with available uncertainties, changing the error bar threshold, and/or adding a range of uncertainties ($0.02\thru0.5 \pp$) in quadratures based on different criteria. The fitted polarimetric parameters change for these optical data; thus, the uncertainties of these parameters were difficult to quantify; however, our main results, discussed in Sect. \ref{s:disc}, were not affected by these variations.

\subsection{Polarimetric parameters}
The final fitting results are presented in Table \ref{tab: res-fit}. The best fit parameter values are the least-square solution, while the error bars are from the 16th and 84th percentiles of the MC samples. The MC median is very close to that of the least-square solutions, differing by only $\ll 1\sigma$. For Ceres, we also included data from \cite{2022PSJ.....3...90M}. More detailed information on the MC simulation is provided in Appendix \ref{s: mcmc}.

\begin{table*}[tb!]
\caption[Fitting Results.]{Fitting results.}
\label{tab: res-fit}
\centering
\begin{tabular}{cc|ccccc}
\hline\hline
Object & Filter & $h$ [$\%/\degr$] & $\ainv$ [$\degr$] & $k$ [$\degr$] & $\amin$ [$\degr$] & $\Pmin$ [\%] \\
\hline
Ceres & J & $0.2570^{+0.0043}_{-0.0042}$ & $17.80^{+0.06}_{-0.05}$ & $6.544^{+0.522}_{-0.435}$ & $6.99^{+0.14}_{-0.13}$ & $-1.754^{+0.025}_{-0.023}$\\
Ceres & H & $0.2567^{+0.0061}_{-0.0047}$ & $17.72^{+0.07}_{-0.06}$ & $7.728^{+0.967}_{-0.628}$ & $7.23^{+0.19}_{-0.14}$ & $-1.641^{+0.031}_{-0.027}$\\
Ceres & \Ks & $0.3008^{+0.0126}_{-0.0064}$ & $17.46^{+0.13}_{-0.07}$ & $13.087^{+6.304}_{-1.567}$ & $7.77^{+0.35}_{-0.14}$ & $-1.636^{+0.057}_{-0.029}$\\
\hline
Vesta & J & $0.0588^{+0.0098}_{-0.0075}$ & $22.72^{+0.65}_{-0.54}$ & $4.124^{+1.124}_{-0.746}$ & $7.05^{+0.64}_{-0.55}$ & $-0.700^{+0.023}_{-0.023}$\\
Vesta & H & $0.0821^{+0.0140}_{-0.0071}$ & $23.26^{+0.41}_{-0.49}$ & $6.283^{+2.391}_{-0.855}$ & $8.38^{+0.70}_{-0.36}$ & $-0.831^{+0.030}_{-0.027}$\\
Vesta & \Ks & $0.0905^{+0.0135}_{-0.0092}$ & $23.96^{+0.58}_{-0.63}$ & $5.067^{+1.175}_{-0.656}$ & $7.92^{+0.51}_{-0.37}$ & $-1.058^{+0.032}_{-0.030}$\\
\hline
\end{tabular}
\end{table*}

Several observations can be made from the abovementioned results: (i) the Ceres NPB does not appear to vary significantly with wavelength, and (ii) Vesta shows a clear trend of changing the NPB with wavelength, including a deepening trend in the $\Ks$ band. The interpretations and discussions of these results are presented in Section \ref{s:disc}.

In addition to the polarimetric parameters, the albedo value of the objects plays a crucial role in understanding their properties. To calculate the albedos of Ceres and Vesta, we utilized the catalog of visible to NIR reflectance spectra\footnote{
\url{http://smass.mit.edu/catalog.php}
}, which covers $\lambda \sim 0.45\thru2.5\um$, along with the filter profile information of the NIC. We performed trapezoidal numerical integration of the spectra using the filter profiles (but ignoring the reflectance and transmission of the dichroic mirrors) and then converted the integrated value back to the physical albedo. This process implicitly assumes that the spectrum used is the same at $\alpha=0\degr$, implying that the backscattering amplitude is assumed to be identical over the wavelength range. The geometric albedo at a wavelength of $\lambda = 0.55\um$ (FC2 filter of the Dawn spacecraft) was adopted as $p_\mathrm{0.55 \um} = 0.094 \pm 0.007$ for Ceres \citep{2017A&A...598A.130C} and $0.38 \pm 0.04$ for Vesta \citep{2013Icar..226.1252L}. The resulting albedo values and their uncertainties are summarized in Table \ref{tab: alb}. For the optical observations, we used a uniform filter profile spanning the wavelength range with $0.01 \um$ bezels (e.g., for $0.45\thru0.60$, wavelengths from $0.45+0.01=0.46 \um$ to $0.60-0.01=0.59 \um$). We used albedos only for $\lambda >0.4\um$.

\begin{table}[!tb]
\caption{Geometric albedos and uncertainties. }
\label{tab: alb}
\centering
\begin{tabular} {cccc}
\hline \hline
Object & Wavelength($\mathrm{\mu m}$)/Filter & Albedo & Uncertainty \\
\hline
Ceres & 0.45\thru0.60 & 0.0931 & 0.0069\\
Ceres & 0.60\thru0.75 & 0.0957 & 0.0071\\
Ceres & 0.75\thru1.00 & 0.0943 & 0.0070\\
Ceres & J & 0.0897 & 0.0067\\
Ceres & H & 0.0921 & 0.0069\\
Ceres & \Ks & 0.0942 & 0.0070\\
\hline
Vesta & 0.45\thru0.60 & 0.372 & 0.039\\
Vesta & 0.60\thru0.75 & 0.403 & 0.042\\
Vesta & 0.75\thru1.00 & 0.322 & 0.034\\
Vesta & J & 0.450 & 0.047\\
Vesta & H & 0.434 & 0.046\\
Vesta & \Ks & 0.404 & 0.043\\
\hline
\end{tabular}
\end{table}

For comparison, the figures in {\PaperI} present Ceres and Vesta overlaid on various parameter spaces along with laboratory samples and other asteroids in the albedo--$\Pmin$--$h$--$\ainv$ space. Among these parameters, the Umov laws ($h$--albedo and $\Pmin$--albedo relations) are shown in Figures 3--5 of \PaperI.
For optical wavelengths, rotational variations in amplitude $<0.1\pp$ and the Umov law were studied for Vesta \citep{2016MNRAS.456..248C}. However, we did not consider the rotational variation.

In the $\Pmin$--$\ainv$ space (Figure 7 in \PaperI), the change in the location of Vesta, likely starting in the \Ks\ band, is noteworthy. Vesta is near S-complex asteroids \citep{2017Icar..284...30B} but moves toward the lunar fine region in the \Ks\ band. For the Moon, most regions exhibit a WD trend (because of the albedo effect and/or $D/\lambda$ effect). A similar WD trend is also observed for Vesta. In contrast, the NPB shows almost no change in Ceres in the $\Pmin$--$\ainv$ space.

\section{Discussion}\label{s:disc}

\subsection{The Umov laws ($h$, $\Pmin$, and albedo)} \label{ss:umow}
The $h$--albedo and $\Pmin$--albedo relations are shown in Figures 3--5 of \PaperI. First, the two objects in the NIR region roughly follow Umov laws. To the best of our knowledge, this is the first report on the location of small airless bodies in these parameter spaces at NIR wavelengths, even though further investigation is needed to verify its generality. Second, Vesta changes its location toward the fine particles (small $D/\lambda$) in both parameter spaces, which supports the existence of fine particles on Vesta. As mentioned, the Ceres $h$ optical data are susceptible to a few noisy data points near the inversion. Ceres also shows a noticeable trend of increasing $h$ (but not $\Pmin$) only in the \Ks\ band, while the albedo and $\Pmin$ remain nearly constant.

\subsection{$\Pmin$--$\ainv$ space}
The NPB of Vesta clearly shows a trend of WD increasing $\lambda$, as shown in, for example, Figure 6 of \PaperI. However, Ceres stays at a fixed location over the investigated wavelengths ($<0.5\um$ to $>2\um$). We therefore discuss the particle sizes on these objects based on these observations.

\subsubsection{Determining Vesta's particle size}\label{ss:vesta particle size}
Vesta has albedo $\gg 10\%$ all over the wavelengths of interest (Table \ref{tab: alb}), which allows us to conclude that WD of the NPB by the albedo effect (\PaperI) is unlikely. Therefore, the likely cause is the effect of $D/\lambda$. We argue that $D/\lambda \sim 5\thru10$ is reached in the H or \Ks\ band ($\lambda \sim 2\um$), so Vesta is likely covered with particles of size $D \gtrsim 10\thru20 \um$. Finer particles ($\sim 1\um$) should not dominate the surface; if they do, that indicates that we must have detected a WD trend (or a narrowing-and-deepening trend) in the UV-to-J band observations.

This particle size is consistent with previous speculations. Most importantly, \cite{1980Icar...43..172L} was one of the first studies to investigate optical NPB to determine the grain size of Vesta. After testing multiple Howardite, eucrite, and diogenite (HED) meteoritic samples, $\lesssim 20\um$ particles were found to be necessary to reproduce the observed $\ainv$ of Vesta.
However, finer particles resulted in too high an albedo\footnote{
They wrote ``$0.27$ to $0.4$ in the green filter.''
} compared to Vesta's albedo known at the time ($0.22$; \citealt{1977Icar...31..185M}; \citealt{1977LPSC....8.1091Z}). Because of this mismatch, coarse particles are required to reduce the albedo. Finally, taking all these factors into account, they concluded that the eucrite Bereba sample of coarse particles ($>76\um$) covered with fine ($<10\um$) particles (denoted Bereba B) can match both the $\ainv$ and albedo at the same time.

However, based on the Dawn mission, Vesta's geometric albedo is updated to $ p_\mathrm{V} = 0.42$ \citep{2013Icar..226.1252L}, and the phase function in the F2 filter (Figure 19(a) in \citealt{2013Icar..226.1252L}) shows that the albedo at $\alpha=5\degr$ is $\gtrsim 0.3$, compared to earlier values $0.22$. Thus, the logic that led \cite{1980Icar...43..172L} to a large fraction of coarse particles should be reconsidered. Instead, the updated data may support a larger fraction of finer particles, which is consistent with our results.

Later, \cite{1994Metic..29..394H} used an independent method (reflectance spectrum analysis) and found that the finest particle size ($<25\um$) of HED powders best reproduced the Vesta $0.3\thru2.6\um$ spectra. Hence, they also concluded that very fine ($<25\um$) particles are necessary to explain Vesta's reflectance spectrum. Recently, \cite{2011Icar..216..640L} reached a similar conclusion that a single, fine-grained Howardite with a size of $<25\um$ suitably fits the \textit{Hubble} Space Telescope, \textit{Swift}, and International Ultraviolet Explorer (IUE) observations. In addition, \cite{2019MNRAS.483.1952M} showed that, using ray-optic simulation calculations, Vesta's surface is dominated ($>75\%$) by fine Howardite particles of size $<25\um$.

Finally, utilizing the thermal conduction modeling algorithm \citep{2022PSJ.....3...47M}, we found two solutions for Vesta, $D \lesssim 10 \um$ and $D\gtrsim 50\um$, both of which require a porosity $\gtrsim 80\%$. Remarkably, this theoretical modeling, independent of polarimetry, yielded results that align quite closely with our findings. The slight disparity between our findings and those of the theoretical thermal model can be attributed to multiple factors, including that polarimetry detects only the topmost layer of the scattering process, whereas thermal modeling probes are significantly deeper (on the order of the thermal skin depth, which varies from the millimeter scale to as high as the meter scale).

Combining the lower bound from our study and the upper bounds from previous works, we propose that Vesta is covered with $D\sim10\thru20 \um$ particles.

\subsubsection{The case of Ceres} \label{ss:case of ceres}
Because Ceres shows no change in the $\Pmin$--$\ainv$ space, one could argue that $D/\lambda \gg 10$ until the \Ks\ band ($D \gg 20\um$). On the other hand, as noted in \PaperI, the narrowing-and-deepening (ND) trend is not clearly detected across the literature, because of the inherent practical difficulty. Moreover, further experiments are needed to investigate when the ND trend stops and how this trend can differ at low albedos ($A\lesssim 10\%$). Therefore, another possible hypothesis is that the fact that $D/\lambda \ll 1\thru2$, even at the shortest wavelength (i.e., $D \ll 0.5\thru1\um$), is the cause of Ceres's behavior.

From the Dawn observations, the wavelength dependence of the single-particle phase function is found \citep{2019Icar..322..144L}. They found that a considerable fraction of $\lesssim 1\um$ particles can explain this dependence. Some regions near young impact craters on Ceres show blue spectra \citep{2017Icar..288..201S}. Later, \cite{2021NatCo..12..274S} showed that these locations are likely to contain hyperporous, foam-like phyllosilicate structures, which effectively act as Rayleigh scatterers with a characteristic filament size $ \ll 1\,\mathrm{\mu m} $ and filament separation $ \gg 1 \,\mathrm{\mu m} $. These can support the small particle scenario.

Utilizing a thermal modeling algorithm similar to Vesta \citep{2022PSJ.....3...47M}, we found two solutions, $D \lesssim 2 \um$ and $D \gtrsim 30\um$, both with porosities $> 80\% $ (cross-checked with MacLennan, E. 2023, priv. comm.). We can reject neither the small nor large particle scenarios. However, we again stress that both possible scenarios from this work and thermal modeling, which are completely independent approaches, expect for surprisingly consistent particle sizes.

Because the NPB behavior of Ceres seems to be typical of C-complex objects \citep{2022PSJ.....3...90M}, further exploration of the change in the NPB for low-albedo materials is key for understanding C-complex objects. Moreover, these findings will naturally deepen our understanding of the thermal behavior of these materials by cross-validating both approaches.

Finally, we note that Ceres's location in the $h$-albedo space is changed in the \Ks\ band. This may indicate the existence of a fine structure. However, as shown in \PaperI, the deviation for low-albedo ($A\lesssim 10\%$) materials has been investigated less than that for higher-albedo materials; thus, further investigations are needed.

\subsection{Revisiting several previous works}
\cite{1987SvAL...13..219B} demonstrated similar trends in $\Pmin$ for both Vesta and (16) Psyche across UBVRI bands, despite differences in their reflectances. A later study showed that S- and M-types generally exhibit the opposite trend to that of the Umov law, which posits an inverse correlation between the polarization degree ($|\Pmin|$) and albedo \citep{2009Icar..199...97B}. This finding suggested that the Umov law is not the sole mechanism influencing the shape of the NPB. This trend aligns with the WD trend, while the trend for low-albedo objects (WD-like as albedo increases; \PaperI) is also possible\footnote{
  \cite{2000Icar..147...94B} shows the geometric albedo in the V band $ A_V(\alpha=0\degr) \sim 0.21 $ and the intensity ratio $I(\alpha=0.3\degr)/I(\alpha=5\degr) = 1.44$ for the 10 S-type samples; thus, $A_V(\alpha=5\degr) < A_V(0\degr) \times I(5\degr)/I(0.3\degr) =15\%$. Hence, they are at the boundary of low and high albedo regimes investigated in \PaperI.
}. This finding highlights the importance of Vesta because it has a sufficiently high albedo (Table \ref{tab: alb}) and allows us to rule out the presence of the albedo effect with a higher confidence.

\cite{2015MNRAS.446L..11B} mentioned that ``the Umov law may be violated'' based on the spectropolarimetry of the Barbarian asteroids. They found that (7) Iris and (599) Luisa showed that $|\polr(\lambda; \alpha)|$ decreased with increasing reflectance, which matches the Umov law, while (236) Honoria showed the opposite trend over $\lambda$. They further discussed that the difference in polarimetric colors between Honoria and Luisa is puzzling because both are Barbarians; over these wavelengths, the NPB becomes deeper for Honoria and shallower for Luisa.

We emphasize that if the ND trend is present on these Barbarian asteroids, this is an expected behavior. Honoria and Luisa were observed at $\alpha = 7.1^\circ$ (where $\polr$ gets deeper with increasing $\lambda$) and $\alpha = 26.9^\circ$ (slightly below $\ainv$, where $\polr$ gets shallower with increasing $\lambda$), respectively. Indeed, \cite{2023PSJ.....4...93M} found that L-types show an ND trend over the wavelength in the NIR region.

We agree with their interpretation that the NPB of L-type plants is affected by fine grains ($D/\lambda <1\thru2$). This example hints that the Umov law should be applied with care; for example, one may use $h$ or $\Pmax$ ($\Pmin$ may be less accurate). It is natural that the inverse Umov law or even the independence of the polarization degree from the wavelength may be observed at certain phase angles. Similarly, the change in NPB for S-types may indicate the presence of smaller particles on them, which may be related to the explanation in \cite{2019PASJ...71..103H}. A possible test observation would include NIR polarimetry of Q-types because they would not show a clear WD trend as S-types if the WD trend was indeed driven by fine particles.

In \cite{2023A&A...677A.146K}, an anticorrelation was found between $\ainv$ and the asteroid diameter of the C-complex. These authors mentioned that this contradicts the fact that larger asteroids may contain finer particles. We hypothesize that this can also be interpreted as the ND trend being observed: the C-complex objects with diameters $d > 100 \km$ already possess fine particles that can induce the ND trend, so $\ainv$ decreases as $D$ decreases (i.e., a decreasing  $\ainv$ increases $d$). We admit that this should be considered a primitive hypothesis because there was no clear ND trend observed for a single object.

\section{Conclusions and future work}
In this work we have shown that NIR polarimetry can be used as a tool for quantifying grain size. We observed two of the largest airless bodies, (1) Ceres and (4) Vesta, which are Dawn mission targets. These airless objects were chosen because they are very likely to possess very fine particles on their surfaces; hence, $D/\lambda$ is expected to be small enough to indicate the change in the NPB over NIR wavelengths. The data were processed with extreme care to achieve the best possible accuracy. Conclusions from this work are summarized as follows:
\begin{enumerate}
\item The Umov laws (slope--albedo and $\Pmin$--albedo correlations) for airless bodies apparently hold at NIR wavelengths. At the longest wavelengths, however, they begin to deviate, which is attributed to the decreased $D/\lambda$.
\item Vesta showed a characteristic change in the NPB when $\lambda$ was increased to the H and \Ks\ bands ($\lambda \gtrsim 2\um$). Combining our data with those of previous studies, we find that the optically significant grain size on Vesta is $\sim 10\thru 20\um$. This finding agrees with that of multiple previous independent reports. To the best of our knowledge, this report is the first quantification of grain size on airless bodies based on NIR polarimetry.
\item For Ceres, there was no significant change in the NPB ($\Pmin$, $\ainv$). This indicates the potential existence of either very fine particles ($D\ll 1\um$) or large particles ($D\gg20\um$).
\item Our empirical estimation of particle sizes using multiwavelength polarimetry strongly aligns with an independent theoretical modeling approach (thermal modeling). This successful congruence between the two methodologies represents a significant cross-validation of the two scientific approaches.
\end{enumerate}

Because Vesta is known to have high heterogeneity in terms of albedo and composition \citep{2012Sci...336..700R}, disk-resolved polarimetry of Vesta near $\amin$ can reveal valuable hints about the parameters that change the NPB.
A change in the polarimetric slope ($h$) of Ceres in the \Ks\ band may provide important information about its surface structure. A further experimental exploration of darker materials, for example using the samples from (101955) Bennu obtained by OSIRIS-REx \citep{2017SSRv..212..925L} or (162173) Ryugu by Hayabusa 2 \citep{2023Sci...379.8671N}, may further advance our understanding of the case of dark C-complex and related objects.

\begin{acknowledgements}
This work was supported by a National Research Foundation of Korea (NRF) grant funded by the Korean government (MEST) (No. 2023R1A2C1006180). YPB would like to thank Eric MacLennan for helpful cross-checking of the grain size calculations and engaging discussions about Ceres. The authors want to thank Joseph Masiero for providing constructive comments as a referee.
\end{acknowledgements}

\bibliographystyle{aa}
\bibliography{ref}


\begin{appendix}
\section{Details of the polarimetry}\label{s: polarimetry detail}
A set of observations contains four exposures at HWP angles of $ \theta = 0^\circ $, $ 45^\circ $, $ 22.5^\circ $, and $ 67.5^\circ $. In this work, we call the first two and the last two the $q$-subset and $u$-subset, as they are used to obtain Stokes $q \coloneqq Q/I$ and $u\coloneqq U/I$, respectively. For each subset, four intensities were measured (the o- and e-ray for each exposure). The measured intensities of the target are $a$ for the o-ray of the first frame, $b$ for the e-ray of the first frame, $c$ for the o-ray of the second frame, and $d$ for the e-ray of the second frame. The Stokes parameter $s$ (either $q$ or $u$) is estimated as
\begin{equation}\label{eq: Stokes s}
  s = \frac{1 - R}{1 + R} ~,
\end{equation}
where
\begin{equation}\label{eq: Stokes R}
  R = \sqrt{ad/bc} ~.
\end{equation}

To obtain $a$ to $d$, we used simple circular aperture photometry since the point-spread function of the NIC is nearly circular.
We always fixed the sky-estimating region as a circular annulus with inner and outer radii of 70 and 90 pixels, respectively. This is determined so that it is large enough to avoid any flux from our object even when the target is the brightest throughout our whole observational period. For some selected cases, by changing this annulus we obtain a result that changes less than the final 1$\sigma$ error bar for the $\polr$ value. The sky value was estimated based on the SExtrator-like algorithm \citep{1996A&AS..117..393B}. First, 3-sigma clipping was performed for a maximum of 5 iterations to reject extreme pixel values. Then, for the sigma-clipped mean, median, and standard deviation ($s_\mathrm{sky}$), the modal value is $m_\mathrm{sky} = (2.5 \times \mathrm{median}) - (1.5 \times \mathrm{mean})$ if $(\mathrm{mean} - \mathrm{median}) / s_\mathrm{sky} \le 0.3$; otherwise, $m_\mathrm{sky} = \mathrm{median}$. This is a slightly modified version of the widely used mode estimation from a unimodal distribution \citep{DoodsonA1917}. Since the size of the final image for each of the o- and e-rays is $130 \times 410$ pixels \citep{2022_SAG_NICpolpy}, the regions outside these regions are regarded as blank pixels in sky estimation.

In addition to $a$ to $d$, we also calculate their uncertainties by
\begin{equation}\label{eq: delta intensity}
  \Delta a = \sqrt{\frac{(1 + \delta_F) (a + m_\mathrm{sky})}{g} + \pi r_\mathrm{ap}^2 \left [ \left (\frac{R_e}{g}\right )^2 + s_\mathrm{sky}^2 \right ]} ~,
\end{equation}
and similar for $b$ to $d$. Here, $\delta_F$ is a factor for coping with the uncertainty in flat fielding, which we fixed to 0.02 (\citealt{1987PASP...99..191S}, and \texttt{DAOPHOT4} uses the default 0.0075, Stetson, P. B. 2023, priv. comm.), $g$ is the gain factor of the detector in electron/ADU \citep{2022_SAG_NICpolpy}, $r_\mathrm{ap}$ is the aperture radius (Sect. \ref{ss: apmode}), and $R_e$ is the readout noise in the electron unit. The S/N is calculated as $\mathrm{S/N}_a = a/\Delta a$, and so on. The uncertainty in Stokes $s$ is
\begin{equation}\label{eq: Stokes s error}
  \Delta s = \frac{R}{(R+1)^2}
    \sqrt{\left (\frac{\Delta a}{a}\right )^2
      + \left (\frac{\Delta b}{b}\right )^2
      + \left (\frac{\Delta c}{c}\right )^2
      + \left (\frac{\Delta d}{d}\right )^2} ~.
\end{equation}

\subsection{Effect of outliers} \label{ss: outliers}
For each subset, we calculate four S/N values: $\{ \mathrm{S/N}_a,\, \mathrm{S/N}_b,\, \mathrm{S/N}_c,\, \mathrm{S/N}_d \}$. Then, the minimum of these is saved as the representative $\mathrm{S/N}$, and the peak-to-peak difference (maximum minus minimum) is called $\mathrm{dS/N}$. Next, we selected sets only if all the following criteria were met:
\begin{enumerate}
\item $\mathrm{S/N}_s > \texttt{SNR\_min}$ for both $s = q$ and $u$
\item $\mathrm{S/N}_s/d\mathrm{S/N}_s > \texttt{SNR\_factor}$ for both $s = q$ and $u$
\item $\Delta s < \texttt{ds\_max}$ for both $s = q$ and $u$.
\end{enumerate}
We selected $\texttt{SNR\_min} = 20$, $\texttt{SNR\_factor}=5$, and $\texttt{ds\_max}=0.5 \%$. When we change these, some data points are affected, but the effects on $\Pmin$ and $\ainv$ are marginal. We introduce this screening method to remove the observations that were made when thin clouds passed through the field of view (i.e., when the S/N was too small or changed rapidly within two consecutive exposures). The choice of these parameters generally does not change the $\polr$ value at all (much smaller than the 1$\sigma$ error bar). For the case of the H band of Vesta on UT 2019-11-08, we found that different combinations of $\texttt{SNR\_min} $ and $\texttt{SNR\_factor}$ (e.g., 30 and 4) can change it by the order of the $0.1\thru0.2 \pp$ level, but this change is still within the order of the error bar. Even if this data point at $\alpha = 4.1\degr$ is very uncertain, it is far from $\amin$ and $\ainv$, so the estimated locations of Vesta in the $\Pmin$--$\ainv$ space are not affected.

After this first screening, we applied the generalized extreme Studentized deviate test (\citealt{Rosner1983GESD}, implemented in \texttt{scikit-posthocs}\footnote{\url{https://github.com/maximtrp/scikit-posthocs}, specifically implemented in this pull request, \url{https://github.com/maximtrp/scikit-posthocs/pull/56}}) to remove an unknown number of outliers in the $q$-$u$ plane. This rejection was based on the distance of the data points with respect to the average point in the $q$-$u$ plane, with a significance level of 0.95. We did not find a strong deviation from the final results due to this process.

\subsection{Effect of aperture modes} \label{ss: apmode}
Changing the centroiding algorithm can change the total flux by a small amount, but this change is much smaller than the change in the sky modal values. Thus, the remaining important factor that can affect the final result is how we select the aperture radius. We tested multiple algorithms to determine the best aperture size for estimating the incident flux.

Considering that the typical full width at half maximum of NIC is approximately 15 pixels, we first calculated the aperture sum using a fixed aperture radius of $r_\mathrm{ap}=25,\, 30,\, 35$ and $40$ pixels (denoted as codes \texttt{25}, \texttt{30}, \texttt{35}, and \texttt{40}). Since the full widths at half maximum clearly vary each night, this results in either a loss of flux (small radius) or a larger scatter (large radius). What makes this more challenging is that we sometimes defocused the image intentionally to avoid saturation. Thus, a single-aperture size is sometimes not appropriate, even over a single night.

Another possible method is to change the aperture size for each subset. Since each exposure results in two images of the target (the o-ray and the e-ray), there are four images. For each of these images, we obtained the modal pixel values as a function of the radial distance from the centroid of our target (for distances of 5 to 60 pixels). The modal values were calculated using an algorithm identical to that used for sky modal value estimation by using a circular annulus of 1 pixel thickness. Then, we selected the maximum radius that has a modal value $ > m_\mathrm{sky} + 1 \times s_\mathrm{sky}$ (denote it $r_\mathrm{ap0}$). If the object is too faint and no such radius exists, we use $r_\mathrm{ap0} = 5$. For the four $r_\mathrm{ap0}$ values obtained in this way, we can choose a fixed $r_\mathrm{ap}$ for all four images using either the minimum or maximum of the $r_\mathrm{ap0}$'s (denoted as code \texttt{min} and \texttt{max}, respectively). Finally, the point spread function of the o- and e-rays is expected to be different for NIC since it uses a beam displacer. Thus, we can also try using separate $r_\mathrm{ap}$ for the o-ray and the e-ray, $r_\mathrm{ap, o}$ and $r_\mathrm{ap, e}$, respectively. $r_\mathrm{ap, o|e}$ can be the smallest among the two $r_\mathrm{ap0, o|e}$ values (denoted as code \texttt{minoe}) or the largest (\texttt{maxoe}).

For example, if we assume the $r_\mathrm{ap0}$ values are 15 and 16 for the first frame and 17 and 18 for the second frame (o- and e-ray, respectively), under our notation $r_\mathrm{ap}$ would be $25$ for the \texttt{25} code, regardless of $r_\mathrm{ap0}$ values; $r_\mathrm{ap} = 15$ for \texttt{min}; $r_\mathrm{ap} = 18$ for \texttt{max}; $r_\mathrm{ap, o|e} = 15|16$ for \texttt{minoe}; and $r_\mathrm{ap, o|e} = 17|18$ for \texttt{maxoe}.

\begin{figure}
  \centering
  \includegraphics[width=1\linewidth]{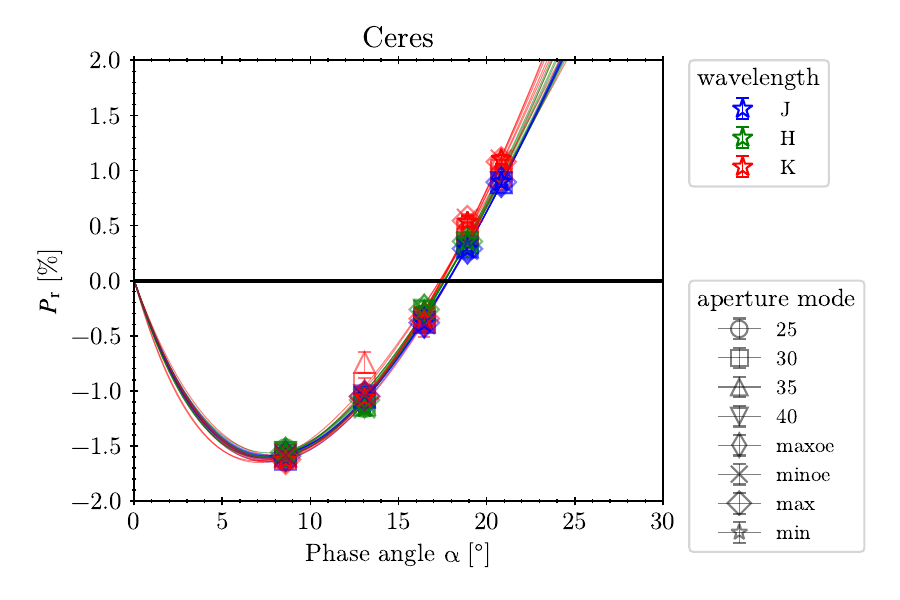}
\caption{Effect of the choice of aperture size in the Ceres case. Different colors correspond to different bands (J, H, and $\mathrm{K_s}$), while different markers correspond to different aperture modes. The solid lines are linear-exponential fits to each band and aperture mode.}
  \label{fig:cerespolalpha-rapmodes}
\end{figure}

\begin{figure}
  \centering
  \includegraphics[width=1\linewidth]{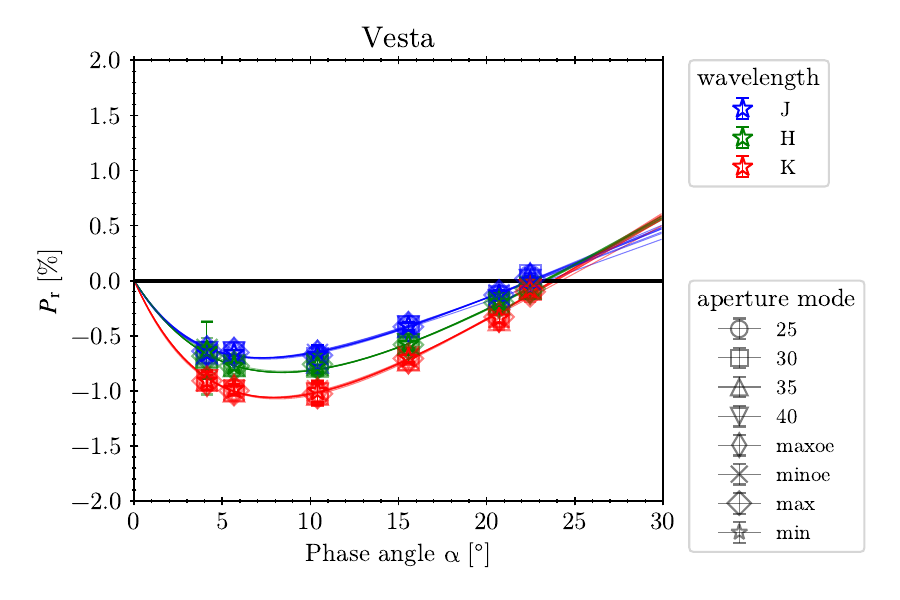}
\caption{Same as Fig. \ref{fig:cerespolalpha-rapmodes} but for Vesta.}
  \label{fig:vestapolalpha-rapmodes}
\end{figure}

We find our final results to be extremely robust to all of these different aperture models (Figs. \ref{fig:cerespolalpha-rapmodes} and \ref{fig:vestapolalpha-rapmodes}). For Vesta, (i) the inversion angle $\ainv$ might be slightly uncertain, especially because there are no data at $\alpha > \ainv$, and (ii) the error bar of the smallest $\alpha$ observation (UT2019-11-08) varies depending on the aperture mode. However, as seen from the figures, this does not change the logic of this paper. All the main results presented in this work are derived using the \texttt{min} aperture model.


\section{Possible systematic uncertainties}\label{ss:systematics}
As usual in any observation or experiment, our observational results may be subject to systematic offsets. Here, we describe possible underlying factors. According to our observations, there are two potential sources of systematic offsets: additive instrumental offsets in terms of the polarization degree and position angle.

An investigation conducted by \cite{TakahashiJ2018SAG} demonstrated that the instrumental polarization degree offset (the so-called instrumental polarization degree) is very small for NIC and is effectively zero within the uncertainty level. The maximum offset occurs in the $\Ks$ band, with values of $(q_\mathrm{off}, u_\mathrm{off}) = (-0.02 \pm 0.30, -0.07 \pm 0.31) \pp$, where the error bars represent the standard deviation of 180 sets of three unpolarized standard stars. For all the other cases, the offsets are $\le 0.03 \pp$. Therefore, the instrumental polarization degree offset is expected to be at most of the same order as the random scatter ($d\polr$). Thus, we decided to skip this offset correction since it is negligible compared to random scatter.

The systematic offset in the instrumental position angle is estimated to be on the order of a few degrees \citep{TakahashiJ2019SAG}. The effect of this offset on $\polr$ is also expected to be very small. According to Eq. (\ref{eq: polr}), an extreme systematic offset of $\Delta \thr = 10\degr$ from $\thr = 0\degr$ would result in only $\Delta \polr/\polr = 0.06$. Since the actual offset is $\Delta \thr \sim 1 \degr$ and our data are $|\polr| < 2 \,\%$, we expect $\Delta \polr \ll 0.1 \pp$ for all our data.

Another potential source of systematic offset comes from the polarization efficiency, $p_\mathrm{eff}$. As mentioned by \cite{TakahashiJ2019SAG}, the NIC has an efficiency of $p_\mathrm{eff} > 0.9$, with values of approximately $100\%$ within the error bars: $(98,\, 95,\, 92) \pm (6, 7, 12)\%$ for the J, H, and $\Ks$ bands, respectively. We applied these efficiency corrections to our data. The uncertainties provided are the standard deviations, not the standard errors. Unlike the previous two offsets discussed, $p_\mathrm{eff}$ is generally considered to have a multiplicative effect \citep{2014SPIE.9147E..4OA}, as it is caused mainly by the imperfect extinction ratio of the polarizer. Therefore, it linearly affects $\polr$ in Eq. (\ref{eq: polr}). This means that $\polr$ might be systematically offset (multiplied) by $\lesssim 0.1 \pp$ in our case (Table \ref{tab: res-pol}). We emphasize, however, that this does not affect our overall logic.

In the work by \cite{2022PSJ.....3...90M}, they also found a hint of instrumental systematic offset with $\Delta P = +0.042$ and $ +0.088 \pp$ in the J and H bands, respectively. In their work, they used an empirical law \citep{1975ApJ...196..261S} with the assumption that the maximum polarization occurs at $0.55\um$ for calibrating the instrumental offsets. Likely stemming from the different calibration processes, our data are slightly offset from theirs (e.g., $0.1\pp$ level near $\amin$ for Ceres). We again emphasize that this difference does not affect our discussion.


\section{Monte Carlo simulation and uncertainties} \label{s: mcmc}
We adopted the Markov chain MC package \texttt{emcee} \cite{2013PASP..125..306F, 2019JOSS....4.1864F} with an affine-invariant ensemble sampler for our fitting process. We first calculated the least-square solution for each object and each filter. The priors for $h$, $\ainv$, and $k$ in Eq. (\ref{eq: linexp}) are chosen to be uniform in the range of $h \in [0.01, 10] $ (in \%/\degr), $\ainv \in [10, 30] $, and $k \in [10^{-5}, 100] $ (both in degrees). The initial positions for 32 chains are randomly selected near this position in the three-dimensional parameter space. Then, with the stretch-move algorithm, 3000 samples per chain are drawn. The first 1000 samples from all the samples were discarded; thus, 64000 MC samples were left per object per filter. The $\amin$ and corresponding $\Pmin$ values are calculated for each sample after the sampling process (Eq. \ref{eq: amin}). We assumed that the uncertainty of each observation was purely Gaussian, so the log-likelihood method used the usual figure of merit:
\begin{equation}
  \mathcal{L} = -\frac{1}{2} \sum_{i=1}^{N}
    \left( \frac{ P_\mathrm{r, obs, i} - \polr(\alpha; h, \ainv, k) }
      {dP_\mathrm{r, obs, i}} \right)^2 ~.
\end{equation}
Here, $P_\mathrm{r, obs, i}$ and $dP_\mathrm{r, obs, i}$ are the observed polarization degree of the $i$th data point and its uncertainty, respectively. The log posterior probability function is calculated by adding the log of the prior function to $\mathcal{L}$.

Then, we discarded samples if they were outside any of the $\pm 5 \sigma$ ranges of the three parameters (the asymmetry $\sigma$ for each parameter was calculated from the 16th, 50th, and 84th percentiles of the MC samples). This process discarded up to a few hundred samples. We checked that the randomness of the MC sampling did not change the results by more than $0.1\sigma$.

To obtain the 1$\sigma$ uncertainty range of the curves in Figs. \ref{fig:ppc} and \ref{fig:ppc+apd}, we first calculated the log-likelihood of each sample and converted it to the usual $\chi^2$ statistic ($\chi^2 = -2 \mathcal{L}$). Then, the samples with
\begin{equation}
    \chi^2 < \mathrm{min}\{ \chi^2 \} + \chi^2_\mathrm{ppf}(0.6827, \nu=3)
\end{equation}
are selected. Here, $\mathrm{min}\{ \chi^2 \}$ is the lowest $\chi^2$ value of all the samples (which should be nearly identical to that of the least-square solution), and $\chi^2_\mathrm{ppf}$ is the percent point function, that is to say, the inverse of the cumulative distribution function of the $\chi^2$ distribution of the degrees of freedom $\nu$. The constant 0.6827 is selected specifically for the ``1$\sigma$'' range. In our case, Eq. (\ref{eq: linexp}) contains three parameters of interest, so $\nu=3$. Model parameters within the 1$\sigma$ uncertainty range were selected in the one-tailed test, which is the method suggested in \citet[][Sect. 15.6]{nr2007}\footnote{\url{http://numerical.recipes/book/book.html}}.

\section{Data and code availability}\label{s:data code avail}
We obtained approximately 8 GB and 13 GB of raw data for Ceres and Vesta, respectively. There are approximately 16 GB of dome flats, and some standard star observations are approximately 16 GB. Then, the preprocessed data (levels 1-4, \citealt{2022_SAG_NICpolpy}) are approximately 3 times the total size (100+ GB of data). Due to the large size of the total dataset, we only uploaded (i) the raw (level 0) data of Ceres and Vesta, (ii) the final (level 4) data of Ceres and Vesta, (iii) the corresponding log files produced from \texttt{NICpolpy} \citep{2022_SAG_NICpolpy}, (iv) the used calibration frames, (v) the automatically generated observational logs in PDF format and (vi) the skymonitor record videos. They are available via Zenodo\footnote{\url{https://doi.org/10.5281/zenodo.7788352}}. Part of the codes and data used for this publication are available via GitHub\footnote{\url{https://github.com/ysBach/BachYP_etal_CeresVesta_NHAO}}

\end{appendix}


\end{document}